\documentclass[runningheads]{llncs}

\usepackage[T1]{fontenc}
\usepackage[utf8]{inputenc}
\usepackage[english]{babel}
\usepackage[kerning]{microtype}
\usepackage{fnpct} %
\usepackage[inline]{enumitem} %
\usepackage{csquotes} %
\usepackage{bold-extra} %
\usepackage{xspace}

\usepackage{nicematrix} %
\usepackage{amsfonts}
\usepackage{amsmath}
\usepackage{amssymb}
\usepackage{mathtools}
\usepackage[T1]{sansmath}

\usepackage{tabularx}
\usepackage{multirow}

\usepackage{algpseudocode}

\usepackage[pdfusetitle, unicode]{hyperref}
\makeatletter
\AtBeginDocument{%
    \def\doi#1{\url{https://doi.org/#1}}}
\makeatother
\usepackage{xspace}
\usepackage{listings}

\AtBeginDocument{%

}

\newcommand{\pr}{\lstinline[mathescape]}
\newcommand{\prc}{\lstinline[language=C]}
\newcommand{\pymwp}{\texttt{pymwp}}
\newcommand{\pymwps}{\texttt{pymwp}{ }} %
\newcommand{\demov}{0.2.1}{}

\newcommand*{\eg}{e.g.\@,\xspace}

\newcommand*{\cf}{cf.\@\xspace}
\newcommand*{\ie}{i.e.\@,\xspace}

\makeatletter
\newcommand*{\etc}{%
    \@ifnextchar{.}%
    {etc}%
    {etc.\@\xspace}%
}
\makeatother

\newcommand{\comm}[1]{\mathtt{#1}}

\definecolor{darkgray}{rgb}{.4,.4,.4}
\definecolor{verbgray}{gray}{0.9}

\lstdefinelanguage{imp}{morekeywords={if,then,else,while,do,loop}}

\lstnewenvironment{console}{\lstset{
    backgroundcolor=\color{verbgray},
    columns=fullflexible,
    basicstyle=\ttfamily\footnotesize,
    showstringspaces=false,
    tabsize=4,keepspaces=true,
 showtabs=true,showspaces=false,
    framesep=2pt, frame=single,
    framerule=.5pt,
    xleftmargin=.01\textwidth,
    xrightmargin=.01\textwidth,
    breaklines=true,postbreak=\mbox{\textcolor{red}{$\hookrightarrow$}\space},
    escapeinside={(*}{*)},
    emph={}
}}{}

\makeatletter
\begin{document}

    \title{\pymwp: A Tool for Guaranteeing Complexity Bounds for C Programs
    \texorpdfstring{%
        \thanks{
            This research is supported by the \href{https://face-foundation.org/transatlantic-study-research/transatlantic-research-partnership/}{Transatlantic Research Partnership}. Rubiano and Seiller are supported by the Île-de-France through the DIM RFSI project \enquote{CoHOp}.
        }}%
    {}
    }
    \titlerunning{\pymwp: A Tool for Guaranteeing Complexity Bounds}
    \author{%
        Clément Aubert\texorpdfstring{\inst{1}\orcidID{0000-0001-6346-3043}}{} \and
        Thomas Rubiano\texorpdfstring{\inst{2}}{} \and
        Neea Rusch\texorpdfstring{\inst{1}\orcidID{0000-0002-7354-5330}}{} \and
        Thomas Seiller\texorpdfstring{\inst{2,3}\orcidID{0000-0001-6313-0898}}{}
    }

    \authorrunning{
    	C. Aubert et al.
    }

    \institute{%
        School of Computer and Cyber Sciences, Augusta University \and
        LIPN---UMR 7030 Université Sorbonne Paris Nord \and
        CNRS
    }

    \maketitle

    \begin{abstract}
        Complexity analysis offers assurance of program's runtime behavior, but large classes of programs remain unanalyzable by existing automated techniques.
        The \emph{mwp}-flow analysis sidesteps many difficulties shared by existing approaches, and offers interesting features, such as compositionality, multivariate bounds, and applicability to non-terminating programs.
        It analyzes resource usage and determines if a program's variables growth rates are no more than polynomially related to their inputs sizes.
        This sound calculus, however, is computationally expensive to manipulate, and provides no feedback if the program does not have polynomial bounds.
        Those two defaults were addressed in a previous work, and prepared for the tool we present here: \pymwp, a static complexity analyzer for \texttt{C} programs based on our improved \emph{mwp}-flow analysis.
    \end{abstract}

    \section{Introduction}
    \label{sec:intro}

    Verifying program's resource usage is particularly important for safety-critical applications:
    if usage exceeds available capacity, program will fail at runtime.
    Although automatic complexity analysis is an active research area, no mainstream tools exist for this task.
    Prior attempts have been successful at \eg analyzing various programming languages and obtaining tight bounds~\cite{giesl2017,hainry2021,Hoffmann2012c}, but are limited in scalability and often lack compositionality~\cite{Carbonneaux2015}.
    Implicit Computational Complexity (ICC) \cite{DalLago2012a} aims at finding syntactic criteria to guarantee program's runtime behavior.
    Since ICC systems offer some of the properties missed by other analysis techniques, it is conjectured that it could bridge the gap in achieving realistic complexity analysis.
    This prompted a series of work~\cite{Aubert2022b,Aubert2022g} that culminates with the tool we present in this paper.
    Our tool is one of the first ICC-inspired applications and the first mechanization of the specific technique it implements.
    It gives early insight of the advancements ICC can provide in automatic complexity analysis and program verification.

    The contribution we present in this paper, {\pymwp}~\cite{pymwp}, is a static analyzer for \texttt{C} programs that computes sound worst-case complexity bounds for final values of input variables.
    It provides a certificate guaranteeing that the program uses throughout its execution at most a polynomial amount of space so that if it terminates, it will do so in polynomial time.
    It offers several useful features not commonly found in alternative complexity analyzers: applicability to non-terminating programs---\eg iteration bounds are
    abstracted---, compositionality, multivariate result, and full automation, because it requires no manual configuration or annotations.
    These features produce fast analysis, detailed feedback, and makes \pymwps suitable for integration in larger compilation toolchains.

    In this paper we demonstrate \pymwps from three perspectives.
    We start with functionality, in \autoref{sec:overview}, with brief theoretical foundations and system design.
    This section gives sufficient background to the theoretical framework to understand how the tool operates,
    computes results, and how to interpret those results.
    We also highlight selected theoretical adjustments and system design approaches that were essential to obtaining a practical application.
    The documentation, \eg at \url{https://statycc.github.io/pymwp/relation/} further explains and exemplifies the modules presented in this section.

    Next, we present a specific implementation challenge related to evaluation of the analysis result.
    This is relevant because the theory underlying \pymwps does not address this potentially exponential-time problem, however it is necessary for implementing the analysis.
    In \autoref{sec:eval} we present the algorithm we developed for \pymwps to solve this problem efficiently.

    Lastly, we discuss user interaction.
    \pymwps is built to support multiple use cases.
    While standard user interaction occurs over command-line interface, the tool can easily be reused and integrated with other systems or pipelined, as we explain in \autoref{sec:use}.
    We envision the developments presented in \pymwps can lead to future improvements in static analysis tools that should be usable in interactive environments (\eg IDEs) with nearly immediate feedback to the programmer.

    \section{Overview of \pymwp}
    \label{sec:overview}

    \subsection{Foundations of the \emph{mwp}-flow analysis -- Briefly}
    \label{sec:mwp}

    The \emph{mwp-flow analysis}~\cite{Jones2009} certifies polynomial bounds on the size of the values manipulated by
    an imperative program. It computes the polynomial bound---if it exists---by computing for each variable a vector tracking how it depends on other variables.
    The vector values are determined by applying the rules of the
    calculus to the commands of the program. A program is assigned a matrix collecting those vectors. While this does
    not ensure termination, it provides a certificate guaranteeing that the program uses throughout its execution at
    most a polynomial amount of space, and as a consequence that if it terminates, it will do so in polynomial time in the size of its inputs.

    Flows characterize controls from one variable to another. In increasing growth rate, they can be of type 0---the
    absence of any dependency---\emph{m}aximum, \emph{w}eak polynomial and \emph{p}olynomial.
    They form a semi-ring~\cite[A.1 and A.2]{Aubert2022i}, and we use $+$ to denote $\max$.
     However, the derivation
    may fail---some programs may not be assigned a matrix---if at least one of the variables used in the body of a loop
    depends \enquote{too strongly} on another, making it impossible to ensure polynomial bounds. 
    In its first declension, the rules of the calculus were non-deterministic: multiple matrices could be assigned to the same program,
    to maximize the opportunities of finding bounds if they existed.
    The derivation stops if no bounds could be inferred, leaving the program only partially analyzed.

	We modified those two latter aspects of the theory~\cite{Aubert2022b}:
    having multiple matrices was space- and time-consuming and
	stopping the derivation as soon as no bound could be found was depriving the programmer from precious feedback.
	As a result, the upgraded analysis now always provides a single matrix capturing all the possible derivations as different \emph{choices} that must be made.
	This required to design the additional mechanism that decides whenever choices not leading to an $\infty$ flow (that represents failure) exist.
	We return to the example below---that uses the dummy condition \prc|while(0)| but could have used an arbitrarily complex condition---throughout this paper and in the demonstration.

    \begin{example}
        \label{ex-theory}

        Analysis assigns to program \pr|foo| the \emph{mwp}-matrix on right.

            { \centering
        \begin{tabular*}{\textwidth}{l @{\extracolsep{\fill}} r}
            \begin{lstlisting}[language=imp]
int foo(int x, int y){
 while(0){x=y+y;}
}
            \end{lstlisting}
            &
            $\begin{pNiceMatrix}[first-row,first-col]
                 & \comm{x} & \comm{y} \\
                 \comm{x} & m + \infty \delta(0, 0) + \infty \delta(1,0) & 0  \\
                 \comm{y} & \infty \delta(0, 0) + \infty \delta(1,0) + w \delta(2,0) & m  \\
            \end{pNiceMatrix}$
        \end{tabular*}}
        The matrix captures the dependencies between variables \pr|x| and \pr|y|, from source variable (row) to target
        variable (column). 
        The complex coefficients in the \pr|x| column express that based on the $0^{\text{th}}$ choice, one will get either $\infty$ and $\infty$ (if the $0^{\text{th}}$ or $1^{\text{st}}$ option is picked) or $m$ and $w$ (for the last option).
        It is easy, here, to see that the only option that yields non-\(\infty\) coefficients is the last option, $(2,0)$.
        
    \end{example}

    \subsection{Concretely implementing the abstract analysis}
    \label{sec:imp} 
    
    The \pymwps tool takes as input a path to a \texttt{C} program, and returns the corresponding matrix and the valid derivation choices if the program passes the analysis, an indication of failure and, if requested, the corresponding matrix if not.
    We decided to implement the tool on a subset of \texttt{C} programming language\footnote{List of supported features:
    \url{https://statycc.github.io/pymwp/features/}}, because it naturally maps to the syntax of the analysis, but the
    technique could be applied similarly to any imperative language.
    The name of the tool alludes to its implementation
    language, Python, selected because of its flexibility and use in previous related
    implementations~\cite{lqicm,Moyen2017,Moyen2017b}.

    \subsubsection{Structures of representation}
    \label{subsec:rep}

    We placed considerable attention in the tool design to choosing suitable data structures. 
    The internalization of the choices \emph{pushes inside} the matrices the non-determinism and introduce the need to decide
    if a series of choices leading to non-\(\infty\) coefficients exist.
    In this section we introduce selected representations and their roles in
    realizing the analysis.

    In \pymwp, the \emph{mwp}-matrix is represented as a \textbf{relation}, whose properties are the input variables of
    the program under analysis, and a matrix collecting variable dependencies.
    We represent the matrix using native lists, with defined basic operations \eg sum, product, resize and fixpoint.
    We omit use of robust matrix libraries to keep the tool dependencies as light as possible.
    Note that since the analysis can produce dense matrices\footnote{For example, \url{https://statycc.github.io/pymwp/demo/\#other_dense.c}.}, it is not possible to use
    algorithms optimized for sparse matrices.

    For each variable pair, the matrix contains a \textbf{polynomial}, an ordered list of ordered monomials. A
    \textbf{monomial} is a pair containing a coefficient value and a list of deltas.
    Each delta captures a choice of derivation rules that internalizes the nondeterminism.
    A \textbf{delta} is a pair $(i, j)$ where $i$ is the value and $j$ is the index in the domain (or, command at which a choice was made).
    To ease the analysis,
    the deltas are sorted--- $\delta(i,j)$ is smaller than $\delta(m,n)$ iff either $j<n$ or $(j=n)$ and
    $(i<m)$--- and no two deltas can have the same index.

    \begin{example}
        \label{ex-poly}

        Consider the dependency flow $\infty \delta(0, 0) + \infty \delta(1,0) + w
        \delta(2,0)$, introduced in
        \autoref{ex-theory}.
        In \pymwp, its representation is 

            {\centering
        \begin{lstlisting}[language=Python,basicstyle=\ttfamily\footnotesize,breaklines=true,postbreak=\mbox{\textcolor{red}{$\hookrightarrow$}\space},]
    Polynomial(Monomial("i",(0,0)), Monomial("i",(1,0)),Monomial("w",(2,0)))
        \end{lstlisting}}
        where coefficients \pr|i| and \pr|w| are the $\infty$- and $w$-flows, respectively, and the tuples
        are the deltas.
        To represent the complete \emph{mwp}-matrix from the example, we would create a relation with 2 variables and a matrix of 4 polynomials.

    \end{example}

   These design decisions allow us to capture \emph{mwp}-matrices in a singular---although complex---relation.
   Performing the analysis then becomes a matter of iteratively mapping commands to vectors, collecting those vectors in matrices,
   then composing the relations.

    \subsubsection{Workflow}

    We describe next, at a high level, the general procedure of performing the \emph{mwp} analysis. Recall, the input is
    a path to a C file. The file may contain multiple functions. Each function is treated as a program under analysis,
    thus we refer to \enquote{program} in the remainder of this description.

    \begin{enumerate}
        \item Parse the input file to obtain an abstract syntax tree (AST).
        \item For each program in the AST:
        \begin{enumerate}
            \item Create an initial relation, $R$, whose matrix is an identity matrix.
            \item Sequentially for each statement in program body:
            \begin{enumerate}
                \item Recursively apply derivation rules to obtain $R_i$.
                \item Compose the $R_i$ with previous relation: $R = R \circ R_i$.
                \item If no valid choice remains\footnote{%
                    We omit the details here; see~\cite[Section 4.4]{Aubert2022b} on how this
                    determination is made.}, terminate analysis.
            \end{enumerate}
            \item Evaluate matrix to find valid derivation choices.
            \item Append to result: (relation $R$, valid choices, success flag).
            The success flag is true when a polynomial bound can be derived and
            false otherwise.
        \end{enumerate}
        \item Return result.
    \end{enumerate}

    \section{Efficiently evaluating the matrix}
    \label{sec:eval}

    The final step of the analysis is the evaluation, which finds a series of choices not leading to \(\infty\)-coefficients.
	While only one matrix needs to be searched, the task is challenging because polynomials can represent arbitrarily complex decision surfaces.
    A naive strategy is to iterate all choices, observe the matrix obtained for each series of choices, and retain those that yield \(\infty\)-free matrices.

    \begin{example}
        \label{ex-slow-eval}

        A naive evaluation of \autoref{ex-theory}'s matrix enumerates all choices, and observe that 
        only one choice, (2,0), produces a matrix without $\infty$ coefficients:

            { \centering
        \begin{tabular*}{\textwidth}{l c @{\extracolsep{\fill}} r c c c}
            & & choice: & (0,0) & (1,0) & (2,0) \\

            $\begin{pNiceMatrix}
                 m + \infty \delta(0, 0) + \infty \delta(1,0) & 0  \\
                 \infty \delta(0, 0) + \infty \delta(1,0) + w \delta(2,0) & m  \\
            \end{pNiceMatrix}$

            & $\Rightarrow$ & &

            $\begin{pNiceMatrix}
                 \infty & 0  \\
                 \infty & m  \\
            \end{pNiceMatrix}$
            & $\begin{pNiceMatrix}
                   \infty & 0  \\
                   \infty & m  \\
            \end{pNiceMatrix}$
            & $\begin{pNiceMatrix}
                   m & 0  \\
                   w & m  \\
            \end{pNiceMatrix}$ \\
        \end{tabular*}}
    \end{example}

    Unfortunately, this approach is exponential.
    For a matrix, whose size depends on the number of variables $V$, and an index $i$ that captures the number of choices introduced during analysis, the complexity is $V^2 \times 3^i$.
    Since variable pairs are represented in the matrix as polynomials containing deltas, the evaluation of both terms is dependent on $i$.
    For larger programs, that introduce many choices, the $i$ increases rapidly, and makes exhaustive search computationally prohibitive.
    A more sophisticated solution was necessary to achieve efficient analysis.

    \subsubsection{Efficient evaluation}
     Our efficient evaluation starts by constructing a set $S$ of all the $\delta$ values attached to an  $\infty$ coefficient present in the matrix.
    The inputs to the procedure are $S$, $i$, and allowed choices \eg $(0,1,2)$ in our case.
    
    \begin{enumerate}[label=\emph{Step \arabic*.}]
\item Simplify $S$ in two ways, iteratively until convergence: replace elements that can be represented by
    a single shorter sequence, then remove supersets.
\item  We now need to negate the remaining elements of $S$, because they represent the choices that yield \(\infty\) coefficients, and the desired output is a representation of valid choices.
	We proceed by initially considering all choices as valid, then eliminating those that lead to failure.
    \begin{enumerate}
        \item Compute the cross product of the remaining elements in $S$.
        \item Create a \emph{choice vector} of size $i$, whose elements represent the allowed choices.
        \item Eliminate those choices that lead to infinity.
        \item Discard invalid and redundant choice vectors.
    \end{enumerate}
    \end{enumerate}
    The result is a disjunction of the remaining choice vectors.

    \begin{example}
        \label{ex-fast-eval}
        An example of an efficient evaluation could be:

\begin{align}
\text{\lstinline[basicstyle=\ttfamily\footnotesize]|S = {((0,0),(2,1),(1,2)),((1,0),(2,1)),((0,0))}|} \tag{Initial \(\delta\)-set}\\
\text{\lstinline[basicstyle=\ttfamily\footnotesize]|S = {((1,0),(2,1),((0,0))}|} \tag{Post-simplification}\\
\text{\lstinline[basicstyle=\ttfamily\footnotesize]|(0,0),(1,0) => [[2],[0,1,2],[0,1,2]]|} \tag{Choice vector 1}\\
\text{\lstinline[basicstyle=\ttfamily\footnotesize]|(0,0),(2,1) => [[1,2],[0,1],[0,1,2]]|} \tag{Choice vector 2}\\
\text{\lstinline[basicstyle=\ttfamily\footnotesize]|[[[2],[0,1,2],[0,1,2]], [[1,2],[0,1],[0,1,2]]]|} \tag{Result}
\end{align}

        To apply the result, we select a choice vector, then choose one value at each vector index. This yields a bounded
        derivation result. The result captures how \eg sequences of choices $(2,0),(0,1),(1,2)$ and $(1,0),(1,1),(2,2)$ are valid. However, any choice
        containing $(0,0)$ is always invalid because it is not allowed by the result. Since this is a positional representation,
        it can be compacted further by omitting the index, \eg sequence $(2,0),(0,1),(1,2)$ equal to $[2 0 1]$.
    \end{example}

    Note that unlike the naive approach, this evaluation is independent of the number of variables or the maximum value of $i$.
    Its efficiency is only concerned with the longest unique sequence of derivation choices that
    leads to infinity. In practice these sequences are short after applying the described simplifications.
    This makes the remaining steps computationally trivial and evaluation result can be obtained nearly instantly.
    It also provides a compact representation of valid choices, even in cases where the representation is arbitrarily complex.

    \section{User Interaction}
    \label{sec:use}

    There are multiple ways to use and interact with \pymwp. It can be used as a standalone command-line tool, or as
    imported Python modules. It can be integrated into other services as a Python package, as we show with the
    \href{https://statycc.github.io/pymwp/demo/}{\pymwps online demo}, which is a web application with \pymwps as a
    package dependency. \pymwps does not modify the input program---it is read-only---so it can be run in parallel or
    independent of other processes. Therefore, it could be integrated into more sophisticated compilation toolchains.
    The development version is available as open-source software~\cite{pymwp}, but the easiest installation is through
    the Python Package Index (PyPI): \texttt{pip install pymwp}. The default interaction command is

\begin{console}
pymwp /path/to/file.c [args]
\end{console}

\noindent where the first positional argument, path to a \texttt{C} file, is required. Optional arguments can be added in
place of \texttt{[args]}. The current list of supported arguments is defined in \texttt{pymwp ---help}. By default,
\pymwps displays a log of debugging information and analysis result on the screen and saves the result to
a file. These default behaviors are customizable by specifying optional arguments.

\begin{example}
    \label{ex-use}

    Analysis of \autoref{ex-theory} program with \pymwp\footnote{\url{https://statycc.github.io/pymwp/demo/\#basics_while_2.c}}.
    Observe that the obtained matrix and available choices match with the original example.

    \vspace{.5em}

\begin{console}
$ pymwp basics/while_2.c
...
MATRIX
------------------------------------------------------------------------
x  |  +m+i.delta(0,0)+i.delta(1,0)  +o
y  |  +i.delta(0,0)+i.delta(1,0)+w.delta(2,0)  +m
------------------------------------------------------------------------
[12:46:00] INFO (analysis): CHOICES: [[[2]]]
...
\end{console}
\end{example}

\clearpage

\bibliographystyle{splncs04}
\bibliography{standalone.bib}

\clearpage

\appendix

\section{Detailed demonstration of \pymwp}
\label{app:sec:demo}

This demo installs \pymwps on the local system,
then uses it to analyze \texttt{C} programs. We assume Unix-like system with \texttt{gcc} and \texttt{wget} installed.
Lines starting with \texttt{\$} are commands to run in a terminal, backslash \texttt{\textbackslash}
is a line wrap for long commands, and other lines are output. Long or irrelevant output is omitted using \texttt{\dots}.

\subsection{Setup}

\subsubsection{Installation}

First check minimum system requirements.
The Python version must be 3.7 or higher, and pip must match Python version.
If the output indicates otherwise, update the system before proceeding.

\begin{console}
$ python3 --version && pip --version
Python 3.10.5
pip 22.2.2 from ../site-packages/pip (python 3.10)
\end{console}

\noindent Next, install the specified version of \pymwps from Python Package Index.

\begin{console}
$ pip install pymwp==(*\demov*)
...
Successfully installed pymwp-(*\demov*)
\end{console}

\noindent Double-check that \pymwps was added to path.

\begin{console}
$ pymwp --version
pymwp (*\demov*)
\end{console}

\noindent This means installation completed successfully.

\subsubsection{Obtain programs for analysis}

We can use \pymwps to analyze any program constructed using the supported \texttt{C} language syntax\footnote{See documentation: \url{https://statycc.github.io/pymwp/features/}.}.
For a quick start, we download a set of suitable programs from the \pymwps development repository.

\vspace{1em}

\noindent Download a \pymwps release as a zip file. It includes the example programs we will analyze during this demo.

\begin{console}
$ wget -O pymwp.zip \
    https://github.com/statycc/pymwp/archive/refs/tags/(*\demov*).zip
\end{console}

\noindent Extract contents of the zip file to a directory, then change to that directory.
For the remainder of this demo, we consider \texttt{pymwp\_demo} as the working directory.

\begin{console}
$ unzip pymwp.zip -d pymwp_demo && cd pymwp_demo
\end{console}

\noindent Copy the example programs to the working directory.

\begin{console}
$ cp -R pymwp-(*\demov*)/c_files/* ./
\end{console}

\noindent Check that example programs were copied successfully.
The programs are categorized into subdirectories and \texttt{readme.md} includes
their descriptions.

\begin{console}
$ ls
basics  implementation_paper  infinite  not_infinite  original_paper  other  pymwp-(*\demov*)  readme.md
\end{console}

\noindent This completes the setup. We are ready to start using \pymwp.

\subsection{Default behavior and arguments}

\pymwps requires one positional argument as input: a path to a \texttt{C} file.
That file is pre-processed using a system \texttt{C} compiler
then converted to an abstract syntax tree.
Debugging information and analysis result are logged to the screen.
The analysis result is also written to a file.
This behavior is customizable to accommodate various runtime scenarios.
For a list of all available options, specify \texttt{---help} flag.

\begin{console}
$ pymwp --help
usage: pymwp [-h] [-o OUT] [--logfile LOGFILE] [--cpp_path CPP_PATH]
             [--cpp_args CPP_ARGS] [--headers HEADERS] [--no_cpp]
             [--no_eval] [--no_save] [-s] [--version] [input_file]

Implementation of MWP analysis on C code in Python.

positional arguments:
  input_file           C source code file to analyze

optional arguments:
  -h, --help           show this help message and exit
  -o OUT, --out OUT    file where to store analysis result
  --cpp_path CPP_PATH  C pre-processor [default: gcc]
  --cpp_args CPP_ARGS  C pre-processor arguments [default: -E]
  --headers HEADERS    C headers dir paths, separate by comma
  --no_cpp             disable C pre-processor
  --no_save            do not write analysis result to a file
  --no_eval            skip evaluation
  --fin                ensure completion even on failure
  --logfile LOGFILE    write console output to a file
  -s, --silent         disable console output
  --version            show program's version number and exit
\end{console}

\subsection{Program analysis}

\subsubsection{Polynomially-bounded examples}

\begin{example}
\label{ex-while2}

Consider a program that contains a \pr|while| loop with a binary operation.
It should seem familiar because it is the same example introduced in the paper (\cf \autoref{ex-theory}).
We re-introduce it in this demo to see it in action.
This program is polynomially bounded in inputs.

\begin{console}
$ cat basics/while_2.c
/*
 * This program tests that a simple while program results in the correct analysis.
 */

int foo(int x, int y){
    while (0) {x = y + y;}
}
\end{console}

\noindent Analyzing \texttt{while\_2.c} with \pymwps we obtain, as expected, a polynomial bound---note that the returned choice is $2$, as discussed in \autoref{ex-theory}.

\begin{console}
$ pymwp basics/while_2.c
...
[12:46:00] INFO (analysis):
MATRIX
------------------------------------------------------------------------
x  |  +m+i.delta(0,0)+i.delta(1,0)  +o
y  |  +i.delta(0,0)+i.delta(1,0)+w.delta(2,0)  +m
------------------------------------------------------------------------
[12:46:00] INFO (analysis): CHOICES: [[[2]]]
[12:46:00] INFO (file_io): saved result in output/while_2.json
[12:46:00] INFO (analysis): Total time: 0.1 s (83 ms)
\end{console}

\end{example}

\begin{example}
\label{ex-notinfty}

Next an example with a conditional statement and 4 input variables.

\begin{console}
$ cat not_infinite/notinfinite_3.c
int foo(int X0, int X1, int X2, int X3){
    if (X1 == 1){
        X1 = X2+X1;
        X2 = X3+X2;
    }
    while(X0<10){
        X0 = X1+X2;
    }
}
\end{console}

\begin{console}
$ pymwp not_infinite/notinfinite_3.c
...
------------------------------------------------------------------------
X0  |  +m+i.delta(0,2)+i.delta(1,2)  +o  +o  +o
X1  |  +p.delta(1,0).delta(2,2)+i.delta(0,2)+i.delta(1,2)+w.delta(2,2)..
X2  |  +i.delta(0,0).delta(1,2)+p.delta(0,0).delta(2,2)+i.delta(1,0)....
X3  |  +i.delta(0,1).delta(0,2)+p.delta(0,1).delta(2,2)+i.delta(1,1)....
------------------------------------------------------------------------
[13:05:50] INFO (analysis): CHOICES: [[[0, 1, 2], [0, 1, 2], [2]]]
...
\end{console}

\noindent Note the increased size of the output matrix (concatenated for brevity).
This is expected: matrix size increases with variable count.
Also observe the \texttt{i}-coefficients ($\infty$) in the matrix: they indicate failure along respective derivation paths.
The output of \texttt{CHOICES} shows that the program is polynomially bounded in inputs, but not all derivation choices yield that bound.
Every derivation choice is valid for the \pr|if| statement, but only one choice is valid for the \pr|while| loop.
For any sequence of choices that meet these constraints, a polynomial bound is guaranteed.

\end{example}

\subsubsection{Examples without polynomial bound}

\begin{example}
\label{ex-exponent}

Consider an exponential program

\begin{console}
$ cat infinite/exponent_1.c
/*
 * This program tests that a simple program computing the
 * exponentiation results in matrix with infinite coefficient in them.
 * Inspired from https://stackoverflow.com/a/213897
 */

int main(int x, int n, int p, int r){
    p = x;
    while (n > 0)
    {
        if (n % 2 == 1)
            r = p * r;
        p = p * p;
        n = n / 2;
    }
}
\end{console}

\noindent We confirm with \pymwps that no polynomial bound exists for this program.

\begin{console}
$ pymwp infinite/exponent_1.c
...
[13:09:03] INFO (analysis): RESULT: main is infinite
[13:09:03] INFO (file_io): saved result in output/exponent_1.json
[13:09:03] INFO (analysis): Total time: 0.1 s (83 ms)
\end{console}

\noindent We can obtain more details about this failure by repeating the analysis with specific instructions to compute the final matrix.
This allows locating the source(s) of failure: here, it originates from variables \texttt{p} and \texttt{r}, \ie the $3^{\text{rd}}$ and $4^{\text{th}}$ rows.

\begin{console}
$ pymwp infinite/exponent_1.c --fin
...
MATRIX
------------------------------------------------------------------------
x  |  +m  +o  +m+i.delta(0,2)+i.delta(1,2)+i.delta(2,2) +i.delta(0,1)...
n  |  +o  +m  +i.delta(0,2)+i.delta(1,2)+i.delta(2,2)  +i.delta(0,1)+...
p  |  +o  +o  +i.delta(0,2)+i.delta(1,2)+i.delta(2,2)  +i.delta(0,1)+...
r  |  +o  +o  +i.delta(0,2)+i.delta(1,2)+i.delta(2,2)  +m+i.delta(0,1)..
------------------------------------------------------------------------
\end{console}

\end{example}

\begin{example}
\label{ex-longer-infty}

Increasing the number of input variables and program statements
makes it difficult to determine if a polynomial bound exists.

\begin{console}
$ cat infinite/infinite_6.c
int foo(int X1, int X2, int X3, int X4){
    if (X3 == 0){
        X1 = X2+X1;
    }
    else{
        X2 = X3+X1;
    }
    while(X4<100){
        X1 = X1+X3;
        X2 = X3+X4;
        X3 = X4+X2;
        X4 = X1+X2;
    }
}
\end{console}

\noindent With \pymwps we can easily determine the result.
We omit the matrix here for brevity, but a detailed analysis, with the \texttt{---fin} flag, reveals that
failure occurs at all program variables inside the \pr|while| loop.

\begin{console}
$ pymwp infinite/infinite_6.c
...
[13:10:19] INFO (analysis): RESULT: foo is infinite
[13:10:19] INFO (file_io): saved result in output/infinite_6.json
[13:10:19] INFO (analysis): Total time: 1.0 s (1035 ms)
\end{console}
\end{example}

\subsubsection{Analysis challenge}

After seeing the examples of programs with and without polynomial bounds, we present the following challenge.
The task is to determine if this program is polynomially bounded in inputs.
Note that it is unknown whether the \pr|while| loop will terminate, however this is not a problem for our analysis.

\begin{console}
$ cat other/dense_loop.c
...
int foo(int X0, int X1, int X2){
    if (X0) {
        X2 = X0 + X1;
    }
    else
    {
        X2 = X2 + X1;
    }
    X0 = X2 + X1;
    X1 = X0 + X2;
    while(X2){X2 = X1 + X0;}
}
\end{console}

\noindent \textbf{Challenge solution} The full matrix must be omitted here for brevity, but
can be inspected as an \href{https://statycc.github.io/pymwp/demo/\#other_dense_loop.c}{online demo}\footnote{\url{https://statycc.github.io/pymwp/demo/\#other_dense_loop.c}}.
The matrix contains following information.

\begin{itemize}
\item For any source variable, where either \texttt{XO} or \texttt{X1} is the target,
the dependencies are consistently polynomially bounded, for all choices.
\item The situation is different between any source variable and \texttt{X2} as the target. Multiple deviation choices fail.
From the matrix, we can also observe that failures occur at the \pr|while| loop, independent of the which branch of the conditional
statement was selected.
\end{itemize}

\noindent There are however multiple sequences of choices that still allow completing the derivation.
The generated choice vector captures the valid choices we can apply to complete the derivation.
Therefore, the solution is yes, the program is polynomially bounded in inputs.

\begin{console}
$ pymwp other/dense_loop.c
...
[23:12:13] INFO (analysis):
MATRIX
------------------------------------------------------------------------
X0  |  +m.delta(0,0).delta(0,2)+p.delta(0,0).delta(1,2)+w.delta(0,0)....
X1  |  +p.delta(0,0).delta(1,2)+p.delta(0,0).delta(2,2)+p.delta(1,0)....
X2  |  +m.delta(0,1).delta(0,2)+p.delta(0,1).delta(1,2)+w.delta(0,1)....
------------------------------------------------------------------------
[23:12:13] INFO (analysis): CHOICES: [[[0, 1, 2], [0, 1, 2], [0, 1, 2], [0, 1, 2], [2]]]
[23:12:13] INFO (file_io): saved result in output/dense_loop.json
[23:12:13] INFO (analysis): Total time: 0.1 s (132 ms)
\end{console}

\subsubsection{Explore further} We suggest reading the \texttt{readme.md} file, in the working directory,
 that describes the examples.
 A formatted version of the readme is
\href{https://statycc.github.io/pymwp/examples/}{available online}\footnote{\url{https://statycc.github.io/pymwp/examples/}}.
 We then suggest analyzing more examples independently.
 After developing sufficient familiarity with \pymwp, create custom programs for analysis.

\subsubsection{Clean up and exit}

\noindent Remove temporary files and examples.

\begin{console}
$ cd .. && rm -rf pymwp.zip pymwp_demo
\end{console}

\noindent If you wish to remove \pymwps from host system, run

\begin{console}
$ pip uninstall -y pymwp
Found existing installation: pymwp (*\demov*)
Uninstalling pymwp-(*\demov*):
  Successfully uninstalled pymwp-(*\demov*)
\end{console}

\noindent This completes the demonstration.

\end{document}